\documentclass{article}

\PassOptionsToPackage{numbers}{natbib}

    \usepackage[preprint]{neurips_2024}

\usepackage[utf8]{inputenc} 
\usepackage[T1]{fontenc}    
\usepackage{hyperref}       
\usepackage{url}            
\usepackage{booktabs}       
\usepackage{amsfonts}       
\usepackage{nicefrac}       
\usepackage{microtype}      
\usepackage{xcolor}         
\usepackage{amssymb}
\usepackage{utfsym}
\usepackage{pdfrender}
\usepackage{tipa}
\usepackage{float}
\usepackage{ulem}
\usepackage{multirow}

\definecolor{mycolor1}{HTML}{720056}
\definecolor{mycolor2}{HTML}{487984}
\definecolor{mycolor3}{HTML}{F5F0ED}
\usepackage{hyperref}
\hypersetup{
    pdfnewwindow=true,      
    colorlinks=true,        
    linkcolor=magenta,  
    citecolor=cyan,  
    filecolor=magenta,      
    urlcolor=purple    
}

\usepackage{amssymb}
\usepackage{graphics}
\usepackage{subcaption}

\newcommand{\newcheckmark}{\raisebox{0.6ex}{\scalebox{0.7}{$\sqrt{}$}}}
\newcommand{\newcrossmark}{\scalebox{0.85}[1]{$\times$}}
\newcommand*{\boldcheckmark}{%
  \textpdfrender{
    TextRenderingMode=FillStroke,
    LineWidth=.6pt, 
  }{\newcheckmark}%
}

\usepackage{todonotes}
\usepackage{xspace}

\title{ManaTTS Persian: a recipe for creating TTS datasets for lower resource languages}

\author{%
  Mahta Fetrat Qharabagh \\
  Department of Computer Engineering \\
  Sharif University of Technology \\
  \texttt{m.fetrat@sharif.edu} \\
  \And
  Zahra Dehghanian \\
  Department of Computer Engineering \\
  Sharif University of Technology \\
  \texttt{zahra.dehghanian97@sharif.edu} \\
  \And
  Hamid R. Rabiee \\
  Department of Computer Engineering \\
  Sharif University of Technology \\
  \texttt{rabiee@sharif.edu} \\
}

\begin{document}

\maketitle

\begin{abstract}

In this study, we introduce ManaTTS, the most extensive publicly accessible single-speaker Persian corpus, and a comprehensive framework for collecting transcribed speech datasets for the Persian language. ManaTTS, released under the open CC-0 license, comprises approximately 86 hours of audio with a sampling rate of 44.1 kHz. Alongside ManaTTS, we also generated the VirgoolInformal dataset to evaluate Persian speech recognition models used for forced alignment, extending over 5 hours of audio. The datasets are supported by a fully transparent, MIT-licensed pipeline, a testament to innovation in the field. It includes unique tools for sentence tokenization, bounded audio segmentation, and a novel forced alignment method. This alignment technique is specifically designed for low-resource languages, addressing a crucial need in the field. With this dataset, we trained a Tacotron2-based TTS model, achieving a Mean Opinion Score (MOS) of 3.76, which is remarkably close to the MOS of 3.86 for the utterances generated by the same vocoder and natural spectrogram, and the MOS of 4.01 for the natural waveform, demonstrating the exceptional quality and effectiveness of the corpus.

\end{abstract}

\section{Introduction}

Text-to-speech conversion has long been an essential task. It is integrated with everyday life, including navigation systems, e-learning, content providing, and much more \cite{googlemaps, speechify-edu, murfai}. But one of the most vital applications of text-to-speech systems is providing accessibility for people with visual impairments, enabling written materials such as electronic device screens to be converted to speech that can be heard rather than read \cite{nvaccess}.

The reason to put an emphasis on the latter application is the lack of open-access high-quality systems. There are actually some Persian text-to-speech models embedded into applications like the Balad map \cite{baladir} and many commercial tools like Narakeet \cite{narakeet}. But there are not any high-quality freely available TTS models that can be used by the more limited audience including the visually impaired and speech domain researchers.
To address these challenges, it is crucial to develop open-access text-to-speech tools, which primarily requires a proper text-to-speech dataset.

An ideal text-to-speech dataset must meet several criteria \cite{naderi2022persian, zen2019libritts}. First, it must
exhibit minimal to no
mismatches in transcripts. Second, it should have no background sound including noise or background music. Third, it should have a high sampling rate (at least 24 kHz) to be useful for modern TTS models. It is also beneficial if the transcripts include exact punctuation to help detect stops and intonations. Additionally, the dataset should be large in terms of both total time duration and word coverage. Therefore, it is important for the data source to be diverse and not limited to a specific domain.

Our investigations show that many existing text-to-speech datasets for the Persian language are not publicly available. On the other hand, there are serious challenges with the available data, the most important being non-open licenses, along with issues such as small size, low quality, and limited domain, which will be discussed in the related works review. Hence, the first step toward open-source and open-access text-to-speech models for the Persian language, and the main focus of the current study, is to prepare such a clean, large-scale and open-source dataset.

In this work, we introduce a new dataset called "ManaTTS". The word Mana means "Enduring" and is derived from the name of a monthly magazine devoted to the blind community, called Nasl-e-Mana \cite{naslemana}, which has been the source of our dataset. The magazine is publicly available, and the content providers were receptive to publishing the dataset with an open CC-0 license.\footnote{The dataset is available at https://huggingface.co/datasets/MahtaFetrat/Mana-TTS} The ManaTTS corpus has the following characteristics:

\begin{itemize}
\item \textbf{Sampling rate:} All the audio files have a sampling rate of 44.1 kHz.
\item \textbf{Speakers:} The entire dataset is recorded by a single female speaker.
\item \textbf{Duration:} It includes 86 hours and 24 minutes of processed and transcribed audio and is by far the largest single-speaker dataset in Persian.
\item \textbf{License:} It is distributed under the open CC-0 1.0 license, enabling educational and commercial use.
\item \textbf{Environment:} The data is mostly recorded in a silent environment and processed to remove potential background music.
\item \textbf{Processing Method:} The entire processing pipeline of the dataset is available, making it fully reproducible. This pipeline introduces a set of open-source useful tools for speech dataset creation including a new sentence tokenization and forced alignment tool.
\item \textbf{Extendable:} The dataset can be easily extended thanks to the monthly growing Nasl-e-Mana magazine and the fully open pipeline.
\item \textbf{Coverage:} The dataset includes 24113 unique words and encompasses a variety of different topic domains.
\item \textbf{Evaluation:} The dataset is used to train a TTS model and has demonstrated effectiveness and high-quality outputs.
\end{itemize}

We have also collected and processed VirgoolInformal, a smaller dataset comprising 5.63 hours of transcribed speech. This dataset is suitable for evaluating ASR models based on Character Error Rate (CER) and is used to prioritize the ASR models in the alignment tool for this work. For more details, refer to Appendix~\ref{sec:evaluation-of-asr-models}.

The rest of the paper is organized as follows. The next section provides a comprehensive review of the available Persian speech datasets. Section~\ref{sec:dataset-preparation} includes a detailed explanation of the data collection and processing methods. Section~\ref{sec:statistics} describes the statistics of the dataset. Section~\ref{sec:experiments} presents the experimental results. The final sections~\ref{sec:discussion} and~\ref{sec:conclusion} summarize the achievements and limitations of this study.

\section{Related Works}
\label{sec:related-works}

Our analysis encompasses various Persian datasets, including text-to-speech (TTS) datasets (Table~\ref{tb:persian-tts}) and other collections featuring speech-text pairs (Table~\ref{tb:appendix-persian-asr}). These include automatic speech recognition (ASR) datasets, audio-visual speech recognition (AVSR) datasets specific to Persian, a dataset for Persian phoneme recognition (PR), a Persian spoken digit recognition (DR) dataset, as well as multilingual datasets that incorporate Persian language components. While the primary focus of this study and the discussions in this section is on Persian speech corpora for text-to-speech systems, we have also included several well-known English TTS speech datasets for a more comprehensive comparison (Table~\ref{tb:appendix-english-corpora}). To see an extended discussion of the related works, please refer to Appendix~\ref{sec:extended-related-works}

\begin{table}
\caption{\textbf{List of Persian text-to-speech corpora.} \textbf{Size} indicates the duration in hours, and \newcrossmark\ in the \textbf{Natural Text} and \textbf{Natural Audio} columns signifies that the text/audio was synthesized.}
  \label{tb:persian-tts}
  
\begin{tabular}{ccccccc}
\toprule
\textbf{Dataset}  & \textbf{Size} & \textbf{Speakers} & \begin{tabular}[c]{@{}c@{}}\textbf{Natural}\\ \textbf{Text}\end{tabular} & \begin{tabular}[c]{@{}c@{}}\textbf{Natural}\\ \textbf{Audio}\end{tabular} & \textbf{Availability} & \textbf{License}\\
\midrule

\textbf{Mana TTS} & $\sim$ \textbf{86} & \textbf{1} & \boldcheckmark & \boldcheckmark & \textbf{Avail.  } &\textbf{ CC-0 1.0} \\

Arman TTS \cite{shamgholi2023armantts} & $\sim$ 9 & 1 & \newcheckmark & \newcheckmark & Not Avail. & Unknown \\
AmerAndish \cite{naderi2022persian}   & 21  & 1   & \newcheckmark      & \newcheckmark       & Not Avail. & Unknown     \\
tts dataset \cite{magnoliasis2024}   & $\sim$ 16       & 1   & \newcheckmark      & \newcheckmark       & Avail.       & Unknown          \\
TTS audio \cite{moradi2024}  & $\sim$26 & 1   & \newcheckmark      & \newcheckmark       & Avail.       & Proprietary  \\
Persian TTS \cite{alister2024}   & +30 & 1   & \newcheckmark      & \newcheckmark       & Not Avail. & Unknown     \\
tts-famale \cite{magnoliasis2024female}   & $\sim$ 30       & 1   & \newcheckmark      & \newcrossmark     & Avail.       & CC-0 1.0          \\
tts-male \cite{magnoliasis2024male}   & $\sim$ 38       & 1   & \newcheckmark      & \newcrossmark     & Avail.       & CC-0 1.0          \\
Persian Speech \cite{persianspeechcorpus}   & $\sim$ 2.5       & 1   & \newcheckmark      & \newcheckmark      & Avail.       & CC BY-NC-SA 4.0          \\
ParsiGoo \cite{kamtera2024}  & $\sim$ 5 & 6   & \newcheckmark      & \newcrossmark     & Avail.       & CC BY-SA 4.0      \\
\begin{tabular}[c]{@{}c@{}}DeepMine\\ Multi-TTS\end{tabular} \cite{adibian4673655deepmine} & 120 & 67 & \newcrossmark & \newcheckmark & Avail. on req. & Unknown\\
\bottomrule
\end{tabular}
\end{table}

\section{Dataset Preparation}
\label{sec:dataset-preparation}

As mentioned earlier, the raw material of our dataset is crawled\footnote{The crawling script and the entire processing source code is available at https://github.com/MahtaFetrat/ManaTTS-Persian-Speech-Dataset} from the website of the Nasl-e-Mana magazine \cite{naslemana} and is published under the CC-0 1.0 license with the consent of its owners.
A female speaker recorded the majority of the audio files, and we manually removed any files not associated with her to ensure the dataset remained single-speaker.
This data was then processed through a pipeline to obtain pairs of speech and transcripts as output. An overview of the entire pipeline is provided in Figure~\ref{fig:data-processing-pipeline}.

The hardware utilized for the entire processing pipeline and model training consisted of a 12th Gen Intel Core i9-12900K CPU with 24 cores and an NVIDIA GeForce RTX 4090 GPU with 24,564 MiB of memory, supporting CUDA version 12.2.

\subsection{Preprocessing}
Prior to processing speech-text pairs, we preprocess audio and text files separately. The audio files, initially in MP3 format, are converted to WAV files. WAV format offers lossless compression, preserving audio quality throughout processing. Additionally, the audio undergoes processing with a source separation tool, namely Spleeter \cite{spleeter2020}, to eliminate any potential background music and retain only the vocals.

\begin{figure}
    \centering
    \begin{subfigure}{0.465\textwidth}
        \includegraphics[width=\linewidth]{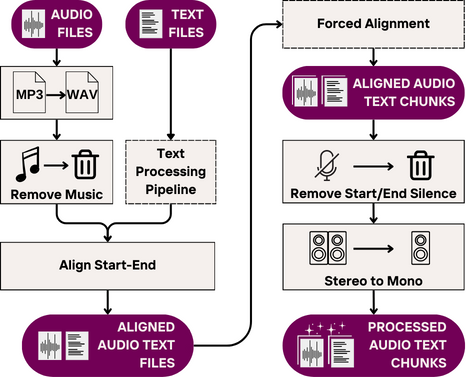}
        \caption{Processing pipeline for audio and text files.}
        \label{fig:data-processing-pipeline}
    \end{subfigure}
    \hfill
    \begin{subfigure}{0.495\textwidth}
        \includegraphics[width=\linewidth]{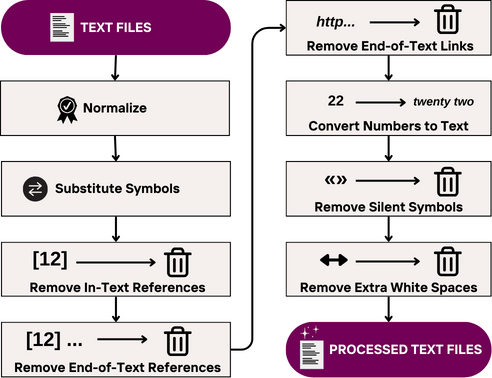}
        \caption{Processing pipeline for text files.}
        \label{fig:text-processing-pipeline}
    \end{subfigure}
    \caption{Dataset processing pipelines.}
    \label{fig:processing-pipelines}
\end{figure}

There is a distinct pipeline for processing the text files, as summarized in Figure~\ref{fig:text-processing-pipeline}. The text data undergoes normalization using Hazm \cite{roshanai} normalizer. This step is crucial as it standardizes the words, ensuring consistency. This simplification reduces the unnecessary details that the TTS model must handle.

The subsequent three steps aim to remove links and references that are typically not meant to be read aloud. These encompass all inline references in the form [NUM], end-of-text references such as author names and book details, and end-of-text links, including entire lines containing URLs.

In the next phase, we addressed the issue of numbers being written differently from how they're spoken. To handle this, we used the parsi-io \cite{parsi.io} tool to detect numbers in the text and convert them into spoken equivalents. Afterward, we trimmed non-essential symbols to streamline the text and decrease the input given to the model. Lastly, we eliminated any extra whitespace, including empty lines, to ready the text for alignment with audio components in later phases, as explained in more detail below.

\subsection{Alignment}
\label{subsec:alignment}

Alignment involves matching audio to its transcript. We have divided this task into two phases. Firstly, we ensure that each audio file contains the same initial and final content as the corresponding text file, which we refer to as start-end alignment. Secondly, we segment the large audio and text files into smaller pieces, typically a few seconds and a few words, in a process known as forced alignment.

Manually performing alignment, especially on large datasets, is an arduous task. Therefore, we opted to automate this process. Our general approach for both alignment phases is rooted in automatic speech recognition (ASR) models. We developed a module that generates reliable hypothesis transcripts for each audio chunk. Subsequently, we attempt to match the hypothesis with the corresponding segment in the ground truth text.

\subsubsection{Transcription Module}
The transcription task can be as straightforward as utilizing a single reliable ASR model to obtain the transcript of a given audio chunk. However, this wasn't our scenario because there is no openly accessible Persian ASR model that is sufficiently reliable to handle this task alone. One common issue with ASR models, for instance, was occasional generation of truncated transcripts for some input cases. Consequently, we chose to integrate multiple ASR models into a transcription module and implement a form of majority voting among them. This approach allows errors from one or two models to be concealed, significantly reducing the likelihood of such defects appearing in the output transcripts.

We utilized five of the top open-access Persian ASR models.
We deliberately selected the ASR models and all tools in our pipeline from the open-source domain, enabling us to publish our work entirely under non-restrictive licenses. A list of the tools used, including the ASR models, can be found in Table~\ref{tb:appendix-tools}.

Some ASR models were accompanied by a reported Word Error Rate (WER). However, they were assessed on different test sets, making them incomparable. To rank and compare the ASR models based on their error rates, we gathered and processed the CC-0 licensed VirgoolInformal dataset, evaluating the models accordingly.\footnote{This dataset is available at https://huggingface.co/datasets/MahtaFetrat/GPTInformal-Persian} Further details regarding this dataset, including its collection method and evaluation results, are provided in the Appendix~\ref{sec:evaluation-of-asr-models}.

\begin{figure}
    \centering
    \includegraphics[width=0.8\linewidth]{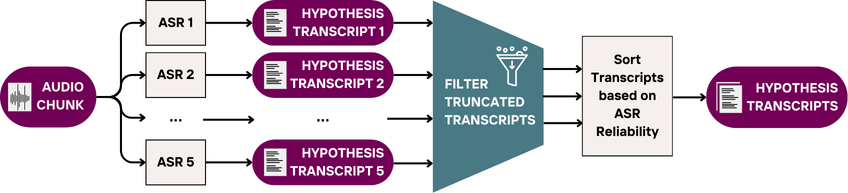}
    \caption{Transcription Module}
    \label{fig:transcription-module}
\end{figure}

The input to the transcription module is a small audio chunk, typically less than 20 seconds in duration. The output comprises a list of eligible transcripts sorted by the reliability of their corresponding ASR models. Figure~\ref{fig:transcription-module} depicts an overview of this module.

The transcripts are generated as follows: initially, a given audio chunk is input into all ASR models. Subsequently, any transcripts shorter than 80\% of the longest is discarded. This step helps address the issue of incomplete transcripts, which was mentioned before. The remaining transcripts that meet the length criteria are then sorted based on the performance of their respective ASR models and returned in a list. Some insightful statistics of this module can be found in Appendix~\ref{sec:transcript-module-statistics}.

\subsubsection{Start/End Alignment}
In the raw dataset, each audio file is paired with a corresponding text file. However, certain factors can cause inconsistencies between the starting/ending points of the audio and its associated text, necessitating start-end alignment processes that may involve removing a few seconds or words from each. The primary factors include:

\begin{itemize}
\item The title and author name read by the speaker, even though not included in the text.
\item Additional resources at the end of the text that are typically unread.
\end{itemize}

The general workflow of the start-end alignment is as follows: the audio is initially segmented based on silent moments, then a search iterates over these segments as potential starting (or ending) points for the audio. For each segment, the most reliable hypothesis transcript is obtained from the transcription module, and the text is searched to find the best matching interval. The pair of trimmed audio and text with the lowest Character Error Rate (CER) between the hypothesis transcript and reference text is selected to determine the start and end of the files.

\subsubsection{Forced Alignment} 
\label{subsubsec:forced-alignment}
The start-end alignment phase generates pairs of audio and text files that are perfectly matched at the beginning and end. However, this format isn't suitable for feeding into a TTS model. The audio and text files must be divided into smaller chunks, typically a few to 15 seconds each. This process is commonly known as forced alignment.

In our search for a forced alignment tool for Persian, we considered Aeneas \cite{aeneas}, which is known for its large community and high performance. However, as noted in their project limitations, "Audio should match the text: large portions of spurious text or audio might produce a wrong sync map." This limitation made Aeneas unsuitable for our needs, as the audio and text could have mismatches due to factors such as:

\begin{itemize}
\item The speaker was provided with a slightly different version of the text to read.
\item The speaker might censor some parts of the text.
\item The speaker might make a mistake and repeat herself to correct it.
\end{itemize}

As a result, we decided to develop our method for forced alignment. The primary workflow of our forced alignment algorithm is illustrated in Figure~\ref{fig:appendix-forced-alignment-flowchart} in Appendix~\ref{sec:supplementary-figures}. 

Initially, the audio is divided into smaller parts using silent intervals. We ensure these audio parts are between 2 and 12 seconds by combining smaller segments or changing the silence detection setting to create smaller parts. The last step is to find a matching section from the reference text. We use the hypothesis transcripts provided by the transcription module until we find a subsection of the text that meets the desired similarity criteria.

The algorithm employs two search methods to find matching text: Interval Search and Gapped Search. Interval Search seeks all sub-strings of the text in the form of \(text[s:i]\) within a defined range. As the name suggests, the Gapped Search would let a missing gap in the text and look for sub-strings of the form \(text[s:j] + text[k:i]\).

The search process halts immediately upon finding a match with \(CER\leq0.05\). Moreover, due to the lower computational cost of Interval Search, matches with \(0.05 < CER \leq 0.2\) at the end of this search are also accepted, avoiding the initiation of the Gapped Search. Suppose neither of the search methods can find a matching substring with \(CER \leq 0.2\). In that case, the process iteratively tests the next hypothesis transcripts until all options are exhausted, resulting in the complete rejection of the chunk.

\subsection{Post-Processing}
In this phase, the audio chunks undergo processing to eliminate any silent segments lasting more than 1 second. It's worth noting that silence removal occurs after forced alignment because silent moments are utilized in segmenting the audio into smaller chunks and should not be removed beforehand. For this task, we utilize the Pydub \cite{pydub} silence module, which is also used for segmenting the audio into chunks. The audio chunks are converted from stereo to mono as the final step.

\section{Statistics}
\label{sec:statistics}

\paragraph{Raw Files:}
 Nasl-e-Mana maintains an archive of 41 magazines spanning over three years. The archive comprises a total of 568 audio-text files. The audio files underwent manual inspection to ensure the dataset consisted of single-speaker recordings. Following this review, four files were excluded from the raw material, resulting in 564 files for automated processing. The duration of the audio files ranged from approximately 0.5 to 34 minutes, with an average duration of about 10 minutes. Similarly, the lengths of the text files varied, with word counts ranging from 44 to 3951 and an average length of 1234 words.

\paragraph{Processed Chunks Count:}
Executing the pipeline on the raw material yielded a minimum, maximum, and average of approximately 4, 398, and 118 chunks per file, respectively, totaling 66172 chunks overall. Roughly 97.98\% of these chunks were automatically accepted as having good quality, while 1338 (about 2.02\%) were rejected due to an unacceptable CER between the hypothesis transcript and the matching text. Consequently, the final dataset comprises 64834 accepted audio-text chunks.

\begin{figure}[htbp]
  \centering
  \begin{minipage}[b]{0.515\textwidth}
    \centering
    \includegraphics[width=\textwidth]{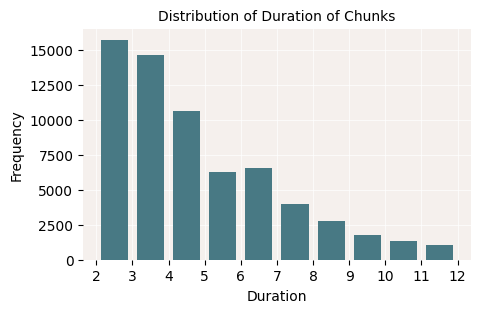}
    \caption{Distribution of the duration of audio chunks.}
    \label{fig:duration-hist}
  \end{minipage}
    \hspace{0.05\textwidth}
  \begin{minipage}[b]{0.385\textwidth}
    \centering
    \includegraphics[width=\textwidth]{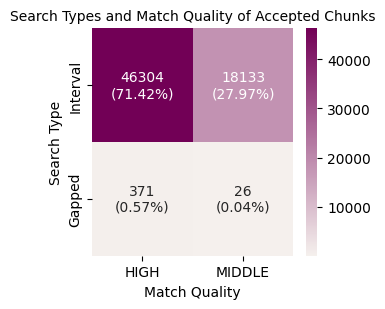}
    \caption{Distribution of search type and match quality of accepted chunks.}
    \label{fig:chunk-distr}
  \end{minipage}
\end{figure}

\paragraph{Accepted Chunks Duration:} As previously mentioned, our pipeline's chunking method guarantees that audio chunks have durations ranging from a minimum of 2 seconds to a maximum of 12 seconds. The histogram depicting the duration distribution of the audio chunks is illustrated in Figure~\ref{fig:duration-hist}.

\paragraph{Accepted chunks search type:}
As outlined in previous sections, two search methods are employed to match a hypothesis transcript with the ground truth text: Interval Search and Gapped Search. It is important to note the preference for the Interval Search method due to its lower computational cost. Consequently, if a match meeting the acceptance criteria is found through Interval Search, further searching with the Gapped Search is unnecessary. Gapped Search is primarily utilized when the ground truth text does not perfectly align with the hypothesis. Analysis of chunk information reveals that approximately 99.39\% of matching text chunks are identified through Interval Search, while the remaining 0.61\% (397 chunks) are the result of Gapped Search.

\paragraph{Accepted chunks match quality:}
As mentioned earlier, there are two threshold levels for CER of audio chunks. The first, labeled HIGH, signifies a perfect match between the hypothesis and ground truth text with a CER less than 0.05. The second, labeled MIDDLE, denotes an acceptable CER ranging from 0.05 to 0.2. Attaining the HIGH threshold during the search prompts an immediate acceptance of the chunk, whereas achieving the MIDDLE threshold would only accept the chunk at the end of each of the search types.

In total, approximately 71.46\% of the chunks (46330 in total) have the HIGH and about 28.54\% of the chunks (18504 in total) have the MIDDLE match quality. Figure~\ref{fig:chunk-distr} illustrates the joint distribution of the search type and match quality of the audio-text chunks, as they are correlated.

\begin{figure}
    \centering
    \includegraphics[width=\linewidth]{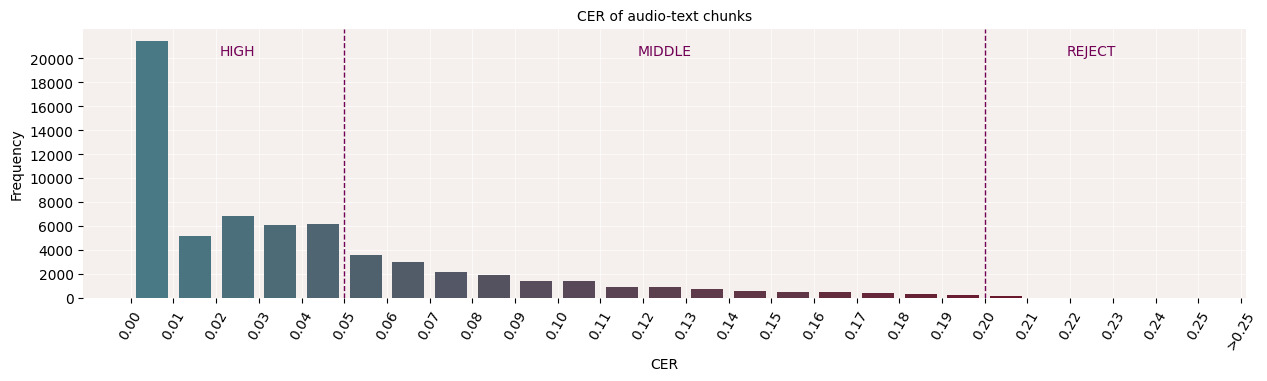}
    \caption{\textbf{Distribution of CER values across all chunks.} The vertical lines denote the threshold values for the HIGH, MIDDLE, and REJECT match qualities as discussed in the section \ref{subsubsec:forced-alignment}.}
    \label{fig:cer-hist}
\end{figure}

It's also intriguing to visualize the distribution of CER values for all the chunks that passed through the dataset creation pipeline. As illustrated in Figure~\ref{fig:cer-hist}, there is only a small number of rejected chunks compared to the matched chunks in the other groups. Manual investigations indicate that these rejected chunks are mostly associated with an underlying discrepancy between the raw audio and text files. Other reasons for rejection include utterances that differ in their written and spoken forms, which will be further discussed in the section \ref{subsec:limitations}.

\begin{figure}
    \centering
    \includegraphics[width=\linewidth]{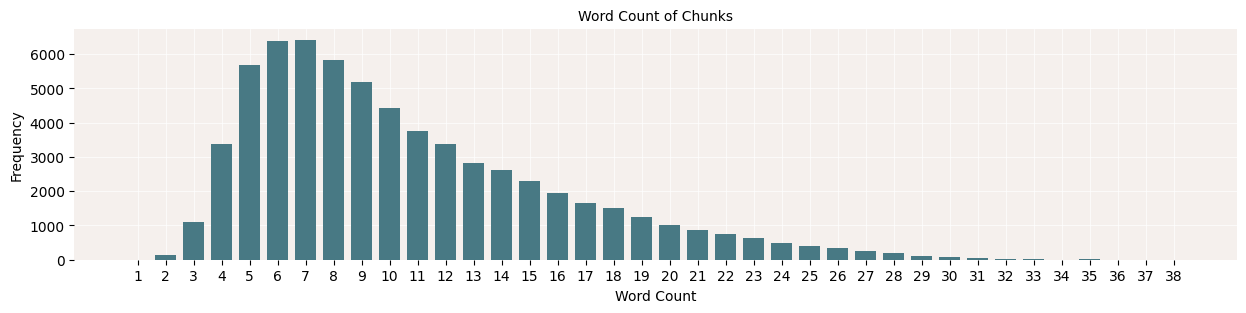}
    \caption{Word count distribution of accepted text chunks.}
    \label{fig:word-hist}
\end{figure}

\paragraph{Word count:}
The accepted chunks exhibit a range of 1 to 38 words, with an average of approximately 11 words per chunk. The histogram illustrating the distribution of word counts can be found in Figure~\ref{fig:word-hist}. Overall, the dataset contains a total of approximately 24,113 unique words.

\section{Experiments}
\label{sec:experiments}

To assess the quality and efficacy of the ManaTTS dataset, we embarked on training a Tacotron2-based TTS model \cite{jia2018transfer} from the ground up using this corpus. This model comprises three main components: First, a speaker encoder, trained on extensive untranscribed datasets, enabling extraction of speaker characteristics from mere seconds of speech. Second, a sequence-to-sequence Tacotron2-based model tasked with converting text into mel-spectrograms, and final component, a vocoder responsible for transforming the mel-spectrogram into the speech waveform.

We adopt a Persian language setup for the text-to-mel-spectrogram module \cite{persian_tacotron2}. The input data undergo resampling to 24 kHz and preprocessing using an FFT size of 2048 and 80 mel-frequency filter banks. We then conduct six training sessions. The learning rate begins at 1e-3 and gradually reduces to 1e-5 in the final session, while the batch size is fixed at 16. With these parameters, the model undergoes training for 320,000 steps and is subsequently used for synthesizing samples for evaluation. \footnote{The full settings and scripts used for training are available at https://github.com/MahtaFetrat/Persian-MultiSpeaker-Tacotron2}

For the speaker encoder and vocoder components, we used the pre-trained encoder and HiFi-GAN \cite{kong2020hifi} vocoder from the previously mentioned work \cite{persian_tacotron2}. HiFi-GAN is trained adversarially, where the generator synthesizes waveforms from spectrograms, and the discriminator distinguishes between synthetic and real waveforms. Due to its non-auto-regressive nature, HiFi-GAN operates faster than earlier vocoders while achieving superior speech quality.

To evaluate the TTS model trained on ManaTTS, we selected five utterances from the latest issue of Nasl-e-Mana magazine that were not included in the training data and synthesized their corresponding speech waveforms using our TTS model. Additionally, we used two open-access Persian TTS models based on VITS \cite{kim2021conditional, kamtera2024persianTTSFemaleVITS} and Glow-TTS \cite{kim2020glow, kamtera2024persianTTSFemaleGlowTTS} as baseline models, generating waveforms for the same utterances for comparison.

Furthermore, we compare the synthesized speech waveforms with the natural speech waveforms of the selected utterances. It is important to note that the vocoder used to generate waveforms from the TTS model's spectrograms was not trained on our dataset. Therefore, to assess the model's spectrogram generation ability fairly, we extracted the mel-spectrograms of the natural utterances and generated their waveforms using the pre-trained vocoder. We then compared these natural spectrogram-generated waveforms to the model's outputs.

We used the subjective Mean Opinion Score (MOS) metric to evaluate our model.\footnote{For other objective methods, please refer to Appendix~\ref{sec:mos-details}.} Scores were collected from 76 native Persian speakers, and the results are presented in Table~\ref{tb:results}. For more details on the evaluation method, please refer to Appendix~\ref{sec:mos-details}.

\begin{table}[]
\caption{\textbf{Subjective assessment of outcomes of the TTS models.} \textbf{GT Spec} refers to the utterances with ground truth spectrograms but HiFi-GAN-synthesized waveforms, and \textbf{GT Waveform} refers to the natural speech samples. The values are presented as mean opinion score (MOS) ± standard deviation (std).}
  \label{tb:results}
  \centering
  
\begin{tabular}{lccccc}
\toprule
\textbf{Source} & VITS & Glow & \textbf{Ours} & GT Spec & GT Waveform\\
\midrule
\textbf{MOS} & \(1.68 \pm 0.80\) & \(1.34 \pm 0.70\) & \(\mathbf{3.76 \pm 1.04}\) & \(3.86 \pm 1.04\) & \(4.01 \pm 1.14\)\\ 
\bottomrule
\end{tabular}
\end{table}

\section{Discussion}
\label{sec:discussion}

The absence of high-quality, open-source/open-access text-to-speech models and datasets for the Persian language has been highlighted in section~\ref{sec:extended-related-works}. Below are some of the critical challenges associated with the available corpora.

\begin{itemize}
    \item The dataset is either inaccessible, lacks a specified license, or is under a restrictive license.
    \item The dataset contains utterances from a limited domain, such as religious contexts exclusively.
    \item The speech is synthesized using a text-to-speech model.
    \item The text is synthesized using a speech-to-text model and has not been adequately verified.
    \item The dataset is limited in size.
\end{itemize}

The ManaTTS dataset introduced in this work is the first Persian language text-to-speech corpus that addresses all the above challenges. This dataset is publicly available under an open CC-0 1.0 license. It includes utterances from a monthly magazine over three years and covers various Persian language utterances. The speech and ground truth text in this dataset are collected by human agents, not synthesized. Most notably, it is the largest single-speaker text-to-speech dataset available to date.

Another notable contribution of this work is that it represents the first fully open-source text-to-speech data collection project for the Persian language. Due to the open code base, all steps, including data crawling, processing, and model training, are reproducible. This approach helps develop additional Persian speech datasets and offers two fundamental benefits.

First, in addition to the standard tools used in typical speech data processing, this project introduces a new sentence tokenization method and a new start-end alignment and forced alignment tool capable of aligning speech and text pairs that are not exact matches but are slightly different. Our work demonstrates that this tool can be effectively used with publicly available Persian ASR models of moderate accuracy, thus contributing yet another open-license tool to the community.

Second, the ever-growing monthly magazine of Nasl-e-Mana and the fully available data collection and processing pipeline make this dataset easily extendable. Future work can obtain an even larger dataset by executing a few scripts.

The experiments reveal that the TTS model, trained on the ManaTTS dataset, achieved a MOS of 3.76, slightly lower than natural spectrograms at 3.86 and natural speech at 4.01. It has even outperformed the ground truth spectrogram and waveform samples in some of the test utterances (refer to Appendix~\ref{sec:mos-details}). This implies that our dataset has acceptable quality and can be used effectively to train Persian text-to-speech models.

\subsection{Limitations}
\label{subsec:limitations}

Despite the contributions of our work, there are some limitations to it. Firstly, although the transcripts are acquired using multiple ASR models and are used to find a match from the ground truth text only if they satisfy certain CER thresholds, this process is not entirely deterministic and is prone to minor errors that might not significantly affect the CER value. Thus, employing superior ASR systems with stricter CER thresholds might further reduce such errors.

Secondly, owing to the pervasive nature of the English language, text in other languages may incorporate English words and phrases. However, our pipeline lacks an explicit mechanism to match these phrases between speech and text. Therefore, if the TTS model is expected to detect and vocalize English subtexts, the pipeline should be modified to include more of these examples in the dataset.

Thirdly, while the pipeline incorporates a mechanism to match the spoken form of numbers, some specifically formatted numeric words remain challenging. For instance, a phone number might be pronounced in various ways, as could date or time. Therefore, a tool capable of converting between these differences in the spoken and written forms of specific numeric data and symbols for the Persian language would be highly beneficial.

Finally, despite the public availability of the raw data, it's important to recognize potential misuse concerns arising from our dataset or processing pipeline, such as voice impersonation. Anonymization emerges as a solution to mitigate these risks, ensuring responsible dataset usage in alignment with privacy and ethical considerations.

\section{Conclusion}
\label{sec:conclusion}

In this work, we proposed a processing pipeline to create a TTS dataset from raw speech and text files. Applying the pipeline to the archive of the Nasl-e-Mana magazine, we published ManaTTS, the largest single-speaker Persian TTS dataset, under the open CC-0 1.0 license. Additionally, we released a smaller transcribed speech dataset, VirgoolInformal, which serves as a valuable test dataset for evaluating ASR models and is utilized in our novel forced alignment method. Evaluating ManaTTS involved training a Tacotron2-based TTS model. The samples synthesized by this model exhibited remarkable naturalness, comparing favorably to both the utterances generated from gold speech spectrograms and natural speech waveforms.

\begin{ack}
We thank the Nasl-e-Mana magazine for their invaluable work and for being so generous with the published dataset license. We also extend our gratitude to the Iran Blind Non-governmental Organization for their support and guidance regarding the need for open access initiatives in this domain.

\textbf{Funding Disclosure:} No funding was received for conducting this study.
\end{ack}

\bibliography{refs}

\newpage
\section*{NeurIPS Paper Checklist}

\begin{enumerate}

\item {\bf Claims}
    \item[] Question: Do the main claims made in the abstract and introduction accurately reflect the paper's contributions and scope?
    \item[] Answer: \answerYes{} 
    \item[] Justification: The Abstract briefly introduces the key findings of our work. The Introduction section highlights the challenges associated with available resources and then enumerates the main characteristics and contributions of our work in nine points.

\item {\bf Limitations}
    \item[] Question: Does the paper discuss the limitations of the work performed by the authors?
    \item[] Answer: \answerYes{} 
    \item[] Justification: There is a Limitations section in the paper introducing four main limitations of the current work and provide suggestions for addressing each.

\item {\bf Theory Assumptions and Proofs}
    \item[] Question: For each theoretical result, does the paper provide the full set of assumptions and a complete (and correct) proof?
    \item[] Answer: \answerNA{} 
    \item[] Justification: The paper does not include theoretical results. 

    \item {\bf Experimental Result Reproducibility}
    \item[] Question: Does the paper fully disclose all the information needed to reproduce the main experimental results of the paper to the extent that it affects the main claims and/or conclusions of the paper (regardless of whether the code and data are provided or not)?
    \item[] Answer: \answerYes{} 
    \item[] Justification: All steps of this study, from the very initial step of data collection scripts to the entire data processing pipeline and evaluation phase, are fully open-source and reproducible. We utilize tools from the open-source domain, and the processed data is also published with the most open license, with consent from its owner. Even the ASR evaluation dataset and processing codes are openly available and reproducible. Additionally, all the new processing tools proposed are also open.

\item {\bf Open access to data and code}
    \item[] Question: Does the paper provide open access to the data and code, with sufficient instructions to faithfully reproduce the main experimental results, as described in supplemental material?
    \item[] Answer: \answerYes{} 
    \item[] Justification: As previously mentioned, all code and scripts are open-source. The raw data source is publicly accessible on the internet, and the processed dataset, suitable for TTS and other speech tasks, is also published under an open CC-0 license.

\item {\bf Experimental Setting/Details}
    \item[] Question: Does the paper specify all the training and test details (e.g., data splits, hyperparameters, how they were chosen, type of optimizer, etc.) necessary to understand the results?
    \item[] Answer: \answerYes{} 
    \item[] Justification: The Data Preparation section outlines the hardware used, while the Experiments section highlights key training parameters. Furthermore, comprehensive details and parameters of the trained model are accessible within the open-source code accompanying this work.

\item {\bf Experiment Statistical Significance}
    \item[] Question: Does the paper report error bars suitably and correctly defined or other appropriate information about the statistical significance of the experiments?
    \item[] Answer: \answerYes{} 
    \item[] Justification: The paper provides error bars (standard deviations) for the subjective assessment of outcomes across various sources in both tabular and graphical representations, offering transparent insights into the variability of the data, as presented in Appendix~\ref{sec:mos-details}.

\item {\bf Experiments Compute Resources}
    \item[] Question: For each experiment, does the paper provide sufficient information on the computer resources (type of compute workers, memory, time of execution) needed to reproduce the experiments?
    \item[] Answer: \answerYes{} 
    \item[] Justification: The paper introduces the hardware specifications utilized for both data processing and experimental model training in the Data Preparation section. 
    
\item {\bf Code Of Ethics}
    \item[] Question: Does the research conducted in the paper conform, in every respect, with the NeurIPS Code of Ethics \url{https://neurips.cc/public/EthicsGuidelines}?
    \item[] Answer: \answerYes{} 
    \item[] Justification: The research aligns with the NeurIPS Code of Ethics, ensuring ethical standards are met throughout the study.

\item {\bf Broader Impacts}
    \item[] Question: Does the paper discuss both potential positive societal impacts and negative societal impacts of the work performed?
    \item[] Answer: \answerYes{} 
    \item[] Justification: The Introduction emphasizes the benefits of a large-scale TTS dataset, particularly its role in providing high-quality models for accessibility purposes. The Limitations section, on the other hand, discusses concerns about potential data misuse.
    
\item {\bf Safeguards}
    \item[] Question: Does the paper describe safeguards that have been put in place for responsible release of data or models that have a high risk for misuse (e.g., pretrained language models, image generators, or scraped datasets)?
    \item[] Answer: \answerYes{} 
    \item[] Justification: We have provided explicit guidance to users against misuse, such as voice impersonation, in dataset utilization. Moreover, the crawled data is sourced from a single reputable website and comprises only text data, reducing potential safety concerns.

\item {\bf Licenses for existing assets}
    \item[] Question: Are the creators or original owners of assets (e.g., code, data, models), used in the paper, properly credited and are the license and terms of use explicitly mentioned and properly respected?
    \item[] Answer: \answerYes{} 
    \item[] Justification: The assets utilized in the paper are explicitly listed in the Supplementary Tables section along with their licenses, and all are sourced from the open-source domain. 
    \item[] Guidelines:
    \begin{itemize}
        \item The answer NA means that the paper does not use existing assets.
        \item The authors should cite the original paper that produced the code package or dataset.
        \item The authors should state which version of the asset is used and, if possible, include a URL.
        \item The name of the license (e.g.,  CC BY 4.0) should be included for each asset.
        \item For scraped data from a particular source (e.g., website), the copyright and terms of service of that source should be provided.
        \item If assets are released, the license, copyright information, and terms of use in the package should be provided. For popular datasets, \url{paperswithcode.com/datasets} has curated licenses for some datasets. Their licensing guide can help determine the license of a dataset.
        \item For existing datasets that are re-packaged, both the original license and the license of the derived asset (if it has changed) should be provided.
        \item If this information is not Avail. online, the authors are encouraged to reach out to the asset's creators.
    \end{itemize}

\item {\bf New Assets}
    \item[] Question: Are new assets introduced in the paper well documented and is the documentation provided alongside the assets?
    \item[] Answer: \answerYes{} 
    \item[] Justification: The code base and tools introduced in this work are thoroughly documented, accompanied by ReadMe files in a publicly accessible repository. All assets are appropriately licensed, with detailed license information provided, including external assets. 

\item {\bf Crowdsourcing and Research with Human Subjects}
    \item[] Question: For crowdsourcing experiments and research with human subjects, does the paper include the full text of instructions given to participants and screenshots, if applicable, as well as details about compensation (if any)? 
    \item[] Answer: \answerYes{} 
    \item[] Justification: Section~\ref{sec:mos-details} contains the instructions provided to human subjects, ensuring transparency in the experimental process. Notably, screenshots are not applicable, and there was no compensation involved.

\item {\bf Institutional Review Board (IRB) Approvals or Equivalent for Research with Human Subjects}
    \item[] Question: Does the paper describe potential risks incurred by study participants, whether such risks were disclosed to the subjects, and whether Institutional Review Board (IRB) approvals (or an equivalent approval/review based on the requirements of your country or institution) were obtained?
    \item[] Answer: \answerYes{} 
    \item[] Justification: As the study did not pose any risks to the subjects, there was no need for risk disclosure. Furthermore, Institutional Review Board (IRB) approvals or equivalent safeguards were not required for this research.

\end{enumerate}


\newpage
\appendix
\section{Extended Discussion of Related Works}
\label{sec:extended-related-works}

Recent advancements in speech recognition and synthesis techniques have led to numerous projects developing systems for the Persian language. Each project comes with its own dataset, each having unique advantages and limitations. This section provides a comprehensive review of Persian speech datasets paired with their corresponding transcripts, detailing their respective merits and drawbacks.

Our analysis encompasses a diverse range of Persian datasets, including text-to-speech (TTS) datasets (Table~\ref{tb:persian-tts}), automatic speech recognition (ASR) datasets, audio-visual speech recognition (AVSR) datasets tailored for Persian, a dataset specifically designed for Persian phoneme recognition (PR), a Persian spoken digit recognition (DR) dataset, and multilingual datasets incorporating Persian language components (Table~\ref{tb:appendix-persian-asr}). While our primary focus and discussions in this section center around Persian speech corpora for text-to-speech systems, we have also included notable English TTS speech datasets for a comprehensive comparison (Table~\ref{tb:appendix-english-corpora}).

\begin{table}[H]

\centering

\caption{\textbf{List of other Persian datasets including speech and text.} The datasets indicated by a plus sign are multilingual, but only the information for the Persian part is shown. \textbf{Size} shows the duration in hours, \textbf{Spks.} stands for the number of speakers, and the columns \textbf{N.T.} and \textbf{N.A.} abbreviate Natural Text and Natural Audio as in Table~\ref{tb:persian-tts}. \textbf{Comm.} stands for a commercial license.}
  \label{tb:appendix-persian-asr}

\begin{tabular}{cccccccc}

\toprule
\textbf{Dataset} & \textbf{Usage} & \textbf{Size} & \textbf{Spks.} & \textbf{N.T}. & \textbf{N.A.} & \textbf{Availability} & \textbf{License}\\
\midrule

DeepMine+ \cite{zeinali2019multi} & ASR & +480 & +1850 & \newcheckmark & \newcheckmark & Paid & Proprietary\\
CMU Wilderness+ \cite{black2019cmu}  & ASR   & 5 & 1   & \newcheckmark      & \newcheckmark       & Avail.       & CC-0 1.0          \\
MLCommons+ \cite{mazumder2021multilingual}  & ASR   & 327 & - & \newcheckmark      & \newcheckmark       & Avail.       & CC BY 4.0        \\
\begin{tabular}[c]{@{}c@{}}Speech \\ Wikimedia\end{tabular}+ \cite{gomez2023speech}  & ASR   & $\sim$ 0.13    & -   & \newcheckmark & \newcheckmark & Avail.       &  CC BY-SA \\
PersianSpeech \cite{persiandataset2024}    & ASR   & $\sim$ 3 & -   & \newcheckmark      & \newcheckmark       & Avail.       & MIT        \\
PersianSpeech \cite{persiandataset2024}   & ASR   & $\sim$ 86       & -   & -       & - & Avail. on req. & MIT        \\
Persian STT \cite{zenodo7486182}   & ASR   & -    & -   & \newcrossmark     & \newcheckmark       & Not Avail.       & CC BY 4.0       \\
Small Farsdat \cite{farsdat}    & ASR   & 5 & 300 & \newcheckmark      & \newcheckmark       & Paid     & \begin{tabular}[c]{@{}c@{}}Non comm. \\ Comm.\end{tabular} \\
Large Farsdat \cite{large-farsdat}   & ASR   & $\sim$ 73       & 100 & \newcheckmark      & \newcheckmark       & Paid     & \begin{tabular}[c]{@{}c@{}}Non comm. \\ Comm. \end{tabular} \\
ASR Farsi \cite{amirpourmand2024}     & ASR   & +300       & -   & \newcheckmark      & \newcheckmark       & Avail.       & CC-0 1.0          \\
CommonVoice+ \cite{commonvoice}  & ASR   & 416 & -   & \newcheckmark      & \newcheckmark       & Avail.       & CC-0 1.0          \\
FarsSpon \cite{asrgooyesh}      & ASR   & +530       & +5300 & \newcheckmark      & \newcheckmark       & Paid     & Proprietary  \\
Shenasa \cite{shenasa2024}   & ASR   & $\sim$300       & -   & \newcheckmark     & \newcheckmark       & Avail.       & GPL-3.0    \\
Persian-SR \cite{persian-speech-recognition}     & ASR   & -    & -   & \newcheckmark      & \newcheckmark       & Avail. on req. & MIT        \\
Arman AV \cite{peymanfard2024multi}   & AVSR  & 220 & 1760  & \newcheckmark      & \newcheckmark       & Avail. on req. & Proprietary  \\
PLRW \cite{peymanfard2022word}   & AVSR  & 30  & 1800  & \newcrossmark     & \newcheckmark       & Not Avail.       & CC BY 4.0         \\

PCVC \cite{malekzadeh2018persian}   & PR    & $\sim$ 1 & 12  & \newcheckmark      & \newcheckmark       & Avail.       & GPL-3.0    \\
PSDR \cite{ralireza2024}   & DR    & -    & -   & \newcheckmark      & \newcheckmark       & Avail.       & GPL-3.0    \\
\bottomrule
\end{tabular}
\end{table}

\begin{table}[H]

\centering

\caption{\textbf{List of well-known English datasets including speech and text.} \textbf{Size} indicates the duration in hours, and \textbf{Non comm.} stands for a non-commercial license.}
  \label{tb:appendix-english-corpora}

\begin{tabular}{cccccc}
\toprule
\textbf{Dataset} & \textbf{Usage} & \textbf{Size} & \textbf{Speakers} & \textbf{Availability} & \textbf{License}\\
\midrule

Hi-Fi TTS \cite{bakhturina2021hi}      & TTS   & 292 & 10  & Avail.       & CC BY 4.0         \\
Libri TTS \cite{zen2019libritts}     & TTS   & 585 & 2456   & Avail.       & CC BY 4.0         \\
BC2013 \cite{blizzard2013}      & TTS   & 300 & 1  & Avail. on req. & Non comm.    \\
VCTK  \cite{edinburgh_datasets}       & TTS   & 44  & 109  & Avail.       & ODC-BY v1.0       \\
LibriSpeech \cite{panayotov2015librispeech}    & ASR   & 982 & 2484  & Avail.       & CC BY 4.0         \\
People's Speech \cite{galvez2021people}         & ASR   & 30k      & -     & Avail.       & CC BY-SA          \\
RyanSpeech \cite{zandie2021ryanspeech}     & ASR/TTS & 10  & 1  & Avail. on req. & CC BY 4.0         \\
LJSpeech \cite{lj_speech}       & ASR/TTS & 24  & 1  & Avail.       & CC-0 1.0          \\
MAILABS \cite{m-ailabs_dataset}        & ASR/TTS & 75  & 2  & Avail.       & BSD \\     
                                    
\bottomrule
\end{tabular}
\end{table}

It is worth noting that the availability of single-speaker TTS datasets in Persian is notably limited compared to their English counterparts. Moreover, the average size of English TTS datasets significantly surpasses that of Persian datasets. This highlights a crucial gap and emphasizes the pressing need for comprehensive, single-speaker Persian datasets to drive progress in research and application development within this domain. In the subsequent part of this section, we will delve into the specifics of each Persian TTS dataset listed in Table~\ref{tb:persian-tts}.

\paragraph{ArmanTTS \cite{shamgholi2023armantts}} is a prominent single-speaker TTS dataset for the Persian language, comprising approximately 9 hours of audio recorded in a professional studio setting at a sampling rate of 22.05 kHz. The audio files are typically about 2.5 seconds long (with a maximum of 12.5 seconds), corresponding to approximately 5 words (and up to a maximum of 30 words), with an average signal-to-noise ratio of 25 dB. Unfortunately, this dataset has not been publicly available yet and the authors have not provided options for access upon request or specified any licensing terms for its use.

\paragraph{AmerAndish \cite{naderi2022persian}} introduced by Naderi et al., is derived from audio books read by a single female speaker. They used a set of automatic tools to read text of PDF files, remove audio noise, and remove audio clips with a different speaker. However, the task of splitting the audio to chunks and matching the chunks with some reference text was performed manually by human agents and later double checked with an ASR system to remove potential mismatches. The resulting dataset includes chunks of 1-12 seconds and  summing up to 21 hours of audio and matching text. Unfortunately, the authors have not provided any means of accessing the dataset or issued a license for its use.

\paragraph{persian tts dataset \cite{magnoliasis2024}} represents another single-speaker Persian resource, featuring approximately 15.6 hours of audio. While the dataset owners have not provided a detailed description, it is evident that the audio is derived from a Persian translation of the Holy Quran. Regrettably, the owners have also not provided any licensing information for the dataset, and it remains unclear whether the audiobook and the corresponding text are free from copyright restrictions. Furthermore, the dataset's exclusive focus on the Holy Quran means it lacks topical and lexical diversity, which is a substantial limitation for developing TTS systems that require a broad range of vocabulary and expressions to perform effectively in varied contexts.

\paragraph{Persian text-to-speech audio \cite{moradi2024}} is a single-speaker corpus with 26 hours of content. This dataset, derived from a Persian translation of the Holy Quran, lacks a detailed description. Similarly to prior datasets, copyright details and licensing status are not provided by the dataset owners, leaving all rights reserved to the original authors.

\paragraph{Persian-text-to-speech \cite{alister2024}} details a Persian TTS model project. Researchers compiled a dataset from over 30 hours of audio sourced from commercially purchased audiobooks, narrated by a female speaker. They segmented the audio into chunks ranging from 3 to 14 seconds using silence detection, then manually aligned these chunks with their corresponding texts. Notably, the purchase of the audiobooks implies copyright restrictions, rendering the dataset non-public and unlicensed.

\paragraph{persian tts dataset (female) \cite{magnoliasis2024female}} is a single-speaker Persian dataset under a CC-0 license, comprising 30 hours of audio synthesized from the Persian text corpus Naab \cite{sabouri2022naab}. The dataset's primary limitation is its fully synthesized audio content, which restricts the performance of TTS models trained on it, as they cannot reach the naturalness of human speech due to the inherent constraints of the used synthesizer.

\paragraph{persian-tts-dataset-male \cite{magnoliasis2024male}} unveils a CC-0 licensed, single-speaker Persian dataset, containing approximately 38 hours of audio. The dataset documentation lacks specifics regarding the source and data collection methodology. However, Initial manual analysis of several audio samples indicates that the content was synthesized using another TTS model.

\paragraph{Persian Speech Corpus \cite{persianspeechcorpus}} presents a Persian TTS dataset, licensed under the Creative Commons Attribution-NonCommercial-ShareAlike 4.0 International License. Featuring recordings from a single male speaker, this corpus encompasses professionally produced studio-quality audio, but totals only 2.5 hours, which may be considered brief for extensive TTS research and development.

\paragraph{ParsiGoo \cite{kamtera2024}} introduces a multi-speaker TTS dataset tailored for the Persian language, secured under the CC BY-SA 4.0 license. This collection comprises about 5 hours of audio, recorded at a sampling rate of 22.05 kHz. The dataset features four distinct speaking styles across six speakers, enhancing its diversity. However, detailed provenance of the audio sources remains unspecified. Manual examination reveals that audio from five speakers is synthesized, while recordings from the sixth speaker are authentically vocal. Regrettably, there is no information provided on the copyright status of these audio files.

\paragraph{DeepMine Multi-TTS \cite{adibian4673655deepmine}} is the first large-scale multi-speaker Persian TTS dataset. It encompasses 120 hours of audio recordings sampled at a rate of 22.05 kHz, featuring contributions from 67 speakers. The dataset primarily consists of audio files obtained from a platform hosting public and freely accessible audio-books. The audio tracks have been processed manually to remove parts that included background music. The transcripts of this dataset were generated using a specific ASR system and then checked manually. The resulting chunks vary in length from 0 to 14 seconds but are mostly between 1 to 10 seconds long. Although this dataset has not been published publicly and lacks a specified license, the authors note that the data will be available on request for only research purposes.


\newpage
\section{Evaluation of ASR Models}
\label{sec:evaluation-of-asr-models}

As detailed in Section~\ref{subsec:alignment}, alignment tools necessitate the ASR models to be arranged based on their reliability. This section elucidates the process undertaken to conduct this assessment.

\subsection{VirgoolInformal Dataset}
To ensure a proper assessment of the ASR models, we required a dataset that had not been seen by these models during their training phase. However, many existing ASR corpora, such as CommonVoice, had been utilized in training these ASRs. Consequently, we opted to create a small, high-quality dataset sourced from a collection of text files for the evaluation process. We deliberately selected informal Persian text,\footnote{Persian language is spoken with slight variations between formal contexts and everyday use.} as it likely contained fewer words familiar to the models. This approach served as a more rigorous test, evaluating the models' ability to accurately transcribe phonemes in audio files from new domains and thus show a low CER.

To gather text for this dataset, we created a tool that distinguishes between formal and informal writing styles. Using this module, we then crawled the Persian blog post website, Virgool \cite{virgool}, and gathered a set of informal text.\footnote{The code for informal text detection is available at https://github.com/MahtaFetrat/Persian-Informal-Text-Detector} A subset of the collected text files was then recorded by a female speaker in a silent environment through different sessions.

The raw files of the dataset comprise 25 pairs of audio and text files from 25 informal-text posts. The total duration of the audio files is approximately 5.63 hours, with a vocabulary of 6840 unique words. The dataset is segmented into smaller audio and text chunks ranging from 2 to 12 seconds, encompassing up to about 24 words each. Similar to the entire study, this dataset is released under a CC-0 license and is accessible to the public.

\subsection{Dataset Processing}

We utilized the pre-processing component as same as our ManaTTS dataset preparation pipeline to obtain clean pairs of audio and text files.
Given that the audio files in this dataset were meticulously recorded from the crawled text files without alteration, they remain an exact match. Consequently, there was no necessity for the start-end alignment process.

This precise correspondence also enables the use of the lighter forced alignment tool, Aeneas. The Aeneas forced alignment tool requires the text files to be split into sentences and then attempts to align the audio with the provided sentences. Therefore, we needed a sentence tokenization tool for the Persian language.

\subsubsection{Sentence Tokenization Method}

The most widely recognized and commonly used tool for this purpose in Persian is the Hazm sentence tokenizer. However, this tool primarily tokenizes based on punctuation, which can result in some very long sentences if the original text is not well-punctuated, a common occurrence in informal text. To address this issue, we integrate a part-of-speech (POS) model into Hazm tokenizer to get a customized sentence tokenization module. This module considers multiple criteria to split the text into more coherent and independent sentences.

The sentence splitting module requires an input minimum and maximum sentence length, along with the input text string. It utilizes the Hazm sentence tokenizer to segment the text into sentences, primarily separated by punctuation marks. Subsequently, it iterates through these sentences, dividing any that exceed the specified maximum length. Conversely, the minimum parameter is employed to avoid excessively short sentence fragments.

To achieve a meaningful split, this module employs the Perpos \cite{bashari2021perpos} POS model to identify verbs within the text. Subsequently, it divides the string around these identified verb positions. Notably, it includes any adjacent symbols and the conjunction word '\textipa{/v\textscripta\textlengthmark/}' (meaning 'and' in English) in the split with the verb. This is because symbols can influence the verb's intonation, and the word '\textipa{/v\textscripta\textlengthmark/}' following the verb is typically phonetically integrated with it and pronounced as '\textipa{/\textschwa\textupsilon/}'.

The Perpos model is also utilized to identify "Ezafe" tags. Words marked with this tag are pronounced in a manner that is linked to the preceding words. Therefore, it is not advisable to split sentences when encountering this tag, as it may result in the audio being interrupted in the middle of the vowel phoneme \textipa{/e/}. To address this consideration during sentence splitting, the module treats a word and all its subsequent Ezafe-tagged words as a single word group while iterating over the text tokens. The complete code for the processing steps of this dataset, including the sentence tokenization module, is publicly available.\footnote{https://github.com/MahtaFetrat/VirgoolInformal-Speech-Dataset}

\subsection{Evaluation}

The audio-text chunks of VirgoolInformal dataset were employed to evaluate and compare the Persian ASR models. Each audio chunk underwent processing through all the ASR models, and the resulting transcripts were recorded. Following a lightweight text processing to eliminate irrelevant symbols and characters from both the hypothesis transcripts and the ground truth text, the CER between these two strings was computed. Subsequently, the average CER of each model on the dataset chunks was taken into account as the performance criterion (See the first column of Table~\ref{tb:asr-models-eval-combined}).

\begin{table}[htbp]
\small
\centering
\caption{Evaluation results of ASR models based on all output transcripts and transcripts after filtering out truncated instances.}
\label{tb:asr-models-eval-combined}
\begin{tabular}{lcc}
\toprule
\textbf{ASR Model} & \textbf{Average CER (All Transcripts)} & \textbf{Average CER (Filtered Transcripts)} \\
\midrule
\textbf{Vosk}   & 0.1128            & \textbf{0.1005} \\
Wav2vec-v3      & \textbf{0.1090}   & 0.1053 \\
Wav2vec-fa      & 0.1411            & 0.1147 \\
Whisper-fa      & 0.1701            & 0.1616 \\
Hezar           & 0.3703            & 0.3715 \\
\bottomrule
\end{tabular}
\end{table}

It's noteworthy that while some ASR models encountered issues with truncated transcripts, they exhibited high-quality transcripts in other instances. Additionally, the transcription module effectively filters out truncated transcripts, alleviating concerns in this regard. These observations led us to first filter out truncated transcripts by excluding those with less than 80\% of the length of the ground truth text. Subsequently, we calculated the average CER of the ASR models based on the remaining outputs. This metric offers insights into the quality of output transcripts independent of truncation issues. The results of this evaluation approach are presented in the second column of Table~\ref{tb:asr-models-eval-combined}.

The first column of Table~\ref{tb:asr-models-eval-combined} also illustrates the ranking of ASR models' reliability utilized in the alignment tools, determined by evaluation results and initial experimental findings. 

Considering metrics from the second column of Table~\ref{tb:asr-models-eval-combined} as the criterion to sort ASR models based on their reliability yields surprising results. As mentioned in the main body of the paper, utilizing the Vosk model as the most reliable ASR resulted in 71.46\% of the chunks being accepted with a HIGH-quality match to the ground truth text. In contrast, if we selected Wav2vec-v3 as the most reliable ASR because of its smaller CER across all the transcripts, this ratio would reduce to 55.23\%. This observation shows that the non-truncated transcripts from the Vosk model were a better match to the ground truth texts, and the second metric reflects this superiority better.


\newpage
\section{Transcript Module Statistics}
\label{sec:transcript-module-statistics}

As detailed in the Transcription Module section, the typical process for aligning text with an audio chunk's transcript unfolds as follows:

\begin{enumerate}
\item The audio chunk undergoes processing by all ASR models, yielding a list of transcripts.
\item Any flawed transcripts, such as those exhibiting repetitive patterns, are identified and eliminated using regular expressions.
\item The longest transcript is singled out, and any transcripts shorter than 80\% of its length are discarded.
\item The remaining transcripts are arranged in order of the reliability rates assigned to the ASR models during evaluation.
\item Sequentially, the transcripts undergo search algorithms until the earliest one meets the designated CER thresholds, at which point the process halts.
\end{enumerate}

\begin{figure}[H]
    \centering
    \includegraphics[width=\linewidth]{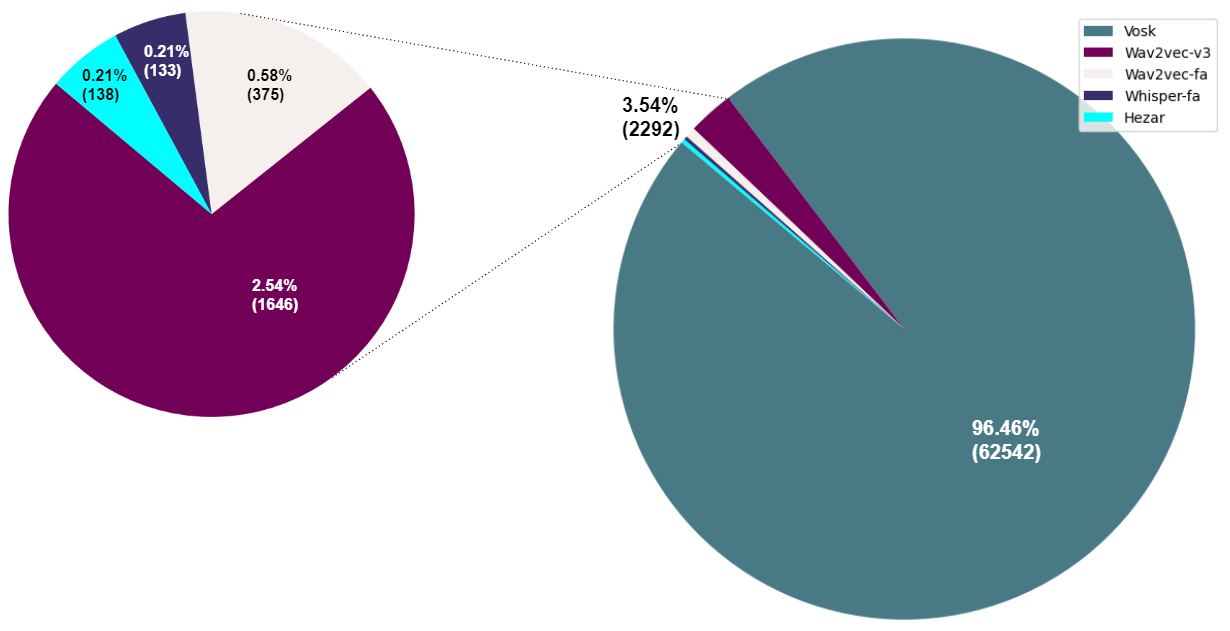}
    \caption{\textbf{Acceptance Ratio by ASR Model.} The values in parentheses represent the exact number of chunks.}
    \label{fig:acceptance-ratio-asr}
\end{figure}

This method naturally leans towards utilizing the most reliable ASR for aligning audio chunks with text. As anticipated, the bulk of the chunks (96.46\%, equating to 62542 chunks) are accepted by the Vosk transcript form. Figure~\ref{fig:acceptance-ratio-asr} illustrates the acceptance ratios of the ASR models, with Vosk contributing the most and Whisper-fa the least.

It's intriguing to observe the effectiveness of the multiple ASR using scheme. This can be explored in two aspects, corresponding to the two strategies employed in the transcription module. Firstly, through majority voting on transcript length, and secondly, by attempting sequential transcript matching until a suitable fit is found.

\begin{figure}[H]
    \centering
    \includegraphics[width=\linewidth]{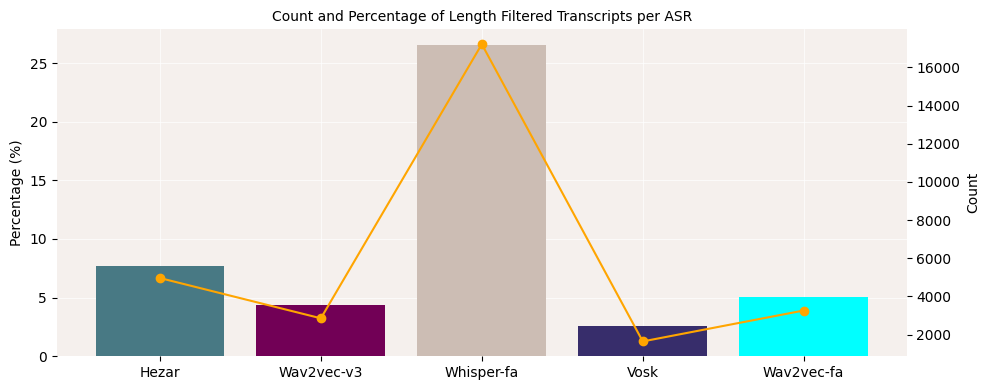}
    \caption{Distribution of transcripts filtered out due to inadequate length.}
    \label{fig:truncated-transcripts-bar-chart}
\end{figure}

Our primary focus lies in assessing the effectiveness of the length majority voting technique in recovering from truncated transcript errors. Upon analyzing our processed data chunks, we observed that transcripts generated by Vosk were excluded from analysis in 1646 audio segments due to their insufficient word count, possibly indicating an error in this particular ASR system. Notably, the truncation error was even more pronounced in less reliable ASRs like Whisper-fa; however, the majority voting technique effectively mitigated its impact. For a visual representation of the number and proportion of rejected transcripts due to the length filter, please refer to Figure~\ref{fig:truncated-transcripts-bar-chart}.

Next, our aim is to assess whether alignment with alternative transcripts contributed to some audio chunks being successfully matched. Table~\ref{tb:transcript-len-disr} presents the number of transcripts that underwent the matching process until an audio chunk was successfully matched with the ground truth text.
It's also worth noting that in 646 of the audio chunks, the transcript from Vosk couldn't be matched to the ground truth text, but it was matched by the transcript from subsequent ASRs.

\begin{table}[H]
\centering
\caption{\textbf{Distribution of transcripts processed to match individual audio chunks.} The table shows the number of chunks aligned using different numbers of transcripts. For example, 64119 chunks were aligned using the transcript from a single ASR, 592 chunks required processing into the second transcript, and so forth.}
\label{tb:transcript-len-disr}
\begin{tabular}{cc}
\toprule
\textbf{Number of Chunks} & \textbf{Number of Processed Transcripts} \\
\midrule
1 & 64119 \\
2 & 592 \\
3 & 74 \\
4 & 39 \\
5 & 10 \\
\bottomrule
\end{tabular}
\end{table}

Our transcription module employs a Majority of Experts (MoE) technique for forced alignment using multiple non-perfect ASR models to mitigate their individual errors. In one experiment, we aimed to assess the robustness of this forced alignment tool by determining how much error of the ASRs it could tolerate.

Before discussing the experiment, it's important to note that regardless of ASR weaknesses, the quality of audio-text chunks from the pipeline remains high. This is because chunks are only accepted if they meet strict CER thresholds, ensuring they uphold a high-quality standard. The primary impact of weaker ASRs is on the number of accepted chunks, not their quality. As ASR errors increase, their transcripts become less similar to the ground truth, resulting in fewer chunks passing the CER thresholds.

To evaluate this, we introduced artificial errors into the ASR outputs, randomly flipping characters in the transcripts. The error rates were uniformly chosen from the ranges $[0, 0.1]$, $[0, 0.2]$, $[0, 0.3]$, $[0, 0.4]$, and $[0, 0.5]$. This produced average CER increases of approximately 
$5\%$, 
$10\%$, 
$15\%$, 
$20\%$, and 
$25\%$ 
respectively. These were substantial increases, especially considering the ASRs already had baseline CERs of $10$-$30\%$ on the VirgoolInformal dataset.

Using these modified ASRs, we performed forced alignment on an audio file that had previously been segmented into 151 chunks without any rejections. The results were impressive, demonstrating the resilience of the MoE approach to ASR degradation. The number of rejected chunks for each error level, as shown in Table~\ref{tb:robustness}, highlights this robustness.

\begin{table}[H]
\centering
\caption{\textbf{Number of rejected chunks at different ASR error levels.} The columns represent the range of additional error introduced into the ASR outputs, while the rows compare the performance of our method (which combines multiple ASRs) with that of using a single ASR. The numbers in the cells indicate how many out of 151 chunks were rejected, with lower values indicating greater robustness and effectiveness.}
\label{tb:robustness}

\begin{tabular}{lccccc}
\toprule
 & $\mathbf{[0, 0.1]}$ & $\mathbf{[0, 0.2]}$ & $\mathbf{[0, 0.3]}$ & $\mathbf{[0, 0.4]}$ & $\mathbf{[0, 0.5]}$ \\
 \midrule
\textbf{Our Method (Multiple ASRs)} & 0 & 2 & 1 & 3 & 9 \\
\textbf{Single ASR} & 1 & 4 & 17 & 53 & 65\\
\bottomrule
\end{tabular}
\end{table}


\newpage
\section{TTS Model Evaluation Details}
\label{sec:mos-details}

In this section, we elaborate on the method used to evaluate the TTS model trained on the MansTTS dataset.

\subsection{MOS Score}

As explained earlier, the model was trained on data from all issues of the Nasl-e-Mana magazine except the most recent one at the time of this study. Thus, the data used to assess the model was selected from this last exclusive issue. We passed the data from this issue through the dataset processing pipeline and obtained the audio-text chunks. We then selected five of these chunks as the evaluation samples.

We then generated the speech waveform of the selected utterances using the following sources:

\begin{enumerate}
\item Baseline Model 1: A VITS-based TTS model trained for the Persian language with an open-access model \cite{kamtera2024persianTTSFemaleVITS}.
\item Baseline Model 2: Another open-access model based on Glow-TTS trained for the Persian language \cite{kamtera2024persianTTSFemaleGlowTTS}.
\item Ours: The model used in our study, which generates spectrograms for a given utterance, and the waveform is generated using a HiFi-GAN vocoder.
\item GT Spec: Waveforms generated from the gold spectrogram of the natural speech using the same HiFi-GAN vocoder as our work.
\end{enumerate}

We accompanied these four classes of utterances with natural audio chunks of the five selected samples and conducted a MOS test using 76 native speakers.

The subjects were prompted as follows: "Rate the voices you hear based on how natural they sound and how likely they are to have been uttered by a human. If you think the voice is completely natural and has no problems, rate it 5. Otherwise, decrease the rating down to 1 based on how robotic it sounds and the problems or noises you notice."

The order of the models for each of the five utterances was shuffled to prevent bias towards consistently rating a specific model the same or being influenced by an increasing/decreasing naturalness trend. They were also not informed which utterances were related to our work or that there was a natural utterance for each sample. The resulting MOS scores can be seen in Table~\ref{tb:utterances} along with the position of the sources in the played samples.

\begin{table}[H]
\caption{\textbf{Subjective assessment of outcomes of the different speech sources per utterance.} \textbf{GT Spec} refers to the utterances with ground truth spectrograms but HiFi-GAN-synthesized waveforms, and \textbf{GT Waveform} refers to the natural speech samples.}
  \label{tb:utterances}
  \centering

  \begin{tabular}{lcccccc}
\toprule
\textbf{} & \textbf{} & \textbf{VITS} & \textbf{Glow} & \textbf{Ours} & \textbf{GT Spec} & \textbf{GT Waveform} \\ \cmidrule{2-7}

\multirow{2}{*}{\textbf{utterance 1}} & \textbf{position} & 3 & 2 & 4 & 1 & 5 \\
 & \textbf{MOS} & 1.78 & 1.44 & 4.24 & 3.69 & \textbf{4.42}\\  \cmidrule{2-7}

\multirow{2}{*}{\textbf{utterance 2}} & \textbf{position} & 5 & 2 & 1 & 4 & 3 \\
 & \textbf{MOS} & 1.65 & 1.26 & \textbf{3.10} & 2.96 & 3.03 \\  \cmidrule{2-7}
 
\multirow{2}{*}{\textbf{utterance 3}} & \textbf{position} & 3 & 2 & 4 & 5 & 1 \\
 & \textbf{MOS} & 1.34 & 1.30 & 3.92 & \textbf{4.20} & 3.96 \\  \cmidrule{2-7}
 
\multirow{2}{*}{\textbf{utterance 4}} & \textbf{position} & 4 & 3 & 1 & 5 & 2 \\
 & \textbf{MOS} & 2.03 & 1.39 & 3.48 & \textbf{4.16} & 4.05 \\  \cmidrule{2-7}
 
\multirow{2}{*}{\textbf{utterance 5}} & \textbf{position} & 4 & 1 & 5 & 3 & 2 \\ 
 & \textbf{MOS} & 1.61 & 1.32 & 4.08 & 4.25 & \textbf{4.57} \\

\bottomrule
\end{tabular}
\end{table}

\begin{figure}[H]
    \centering
    \includegraphics[width=0.7\linewidth]{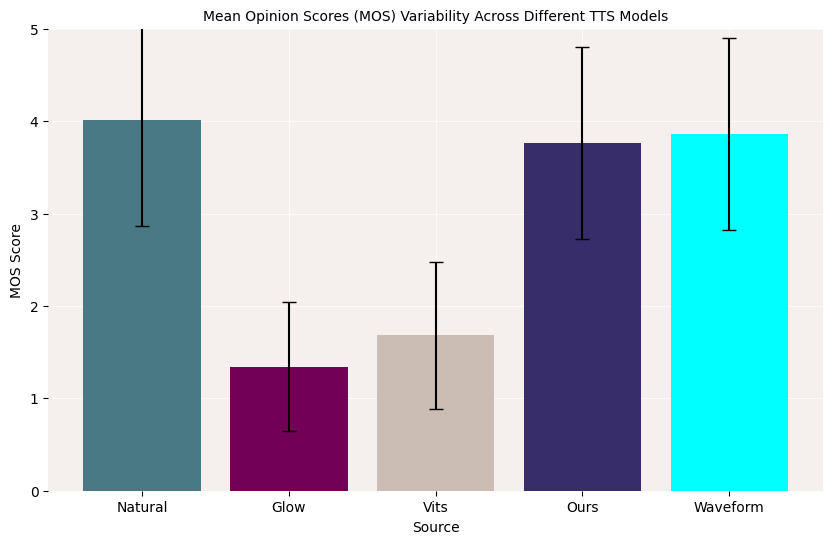}
    \caption{MOS of different sources and their variability.}
    \label{fig:results-std}
\end{figure}

The MOS across all samples and utterances, along with their variability, are presented in Table~\ref{tb:results} and visualized in Figure~\ref{fig:results-std}. The standard deviation of the scores was calculated using the numpy std function on the aggregated scores from all samples of the five utterances. The primary sources of variation in the scores are as follows:

\begin{itemize}
\item Specific sources may appear more natural in some utterances and perform worse in others.
\item Subjects have varying understandings and expectations regarding the naturalness of a speech sample.
\item The random shuffling of the sources' order in each utterance affects the scores given by subjects due to the relative naturalness of the different sources.
\end{itemize}

\begin{figure}[H]
    \centering
    \includegraphics[width=0.6\linewidth]{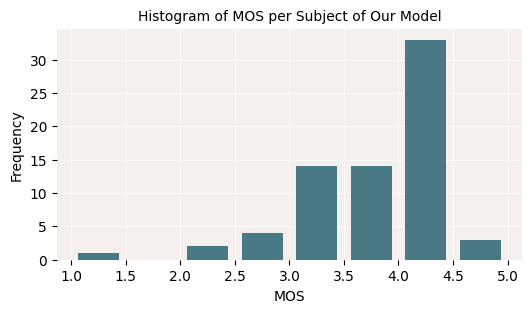}
    \caption{Distribution of Mean Opinion Score (MOS) ratings given to our model by individual subjects.}
    \label{fig:mos-per-subject}
\end{figure}

In addition to examining the model's overall performance, it is insightful to analyze the distribution of Mean Opinion Score (MOS) ratings given to our model by individual subjects. This provides valuable insight into how opinions varied among respondents regarding our model. Figure~\ref{fig:mos-per-subject} presents this distribution, shedding light on how the average scores assigned to our model are spread among respondents.

\subsection{Objective Scores}
In addition to the subjective MOS score, we conducted a more comprehensive evaluation of the trained TTS model using several objective methods. We selected a subset of 100 audio and text chunks, generating audio from these text chunks using our TTS model and two baseline models (VITS and Glow). Additionally, we regenerated the audio from their spectrograms using the vocoder employed by our TTS system. The evaluation metrics included PESQ \cite{pesq}, MCD \cite{mcd}, and APTD (Average Predicted Time Difference in seconds). The results of these metrics are presented in Table~\ref{tb:objective}.

\begin{table}[H]
\centering
\caption{\textbf{Objective assessment of outcomes of the TTS models.} \textbf{GT Spec} refers to the utterances
with ground truth spectrograms but HiFi-GAN-synthesized waveforms.}
\label{tb:objective}

\begin{tabular}{lcccc}
\toprule
 & \textbf{Baseline 1 (Vits)} & \textbf{Baseline 2 (Glow)} & \textbf{Ours} & \textbf{GT Spec} \\
 \midrule
\textbf{PESQ} & 1.05 & 1.06 & 1.11 & 2.89 \\
\textbf{APTD} & 1.4185 & 0.2781 & 0.5783 & 0.0064 \\
\textbf{MCD} & 15.1611 & 18.5218 & 18.5682 & 7.1069\\
\bottomrule
\end{tabular}
\end{table}

We also evaluated intelligibility using two ASR models: 1) Google Speech Recognition API \cite{zhang2017speech}, and 2) Vosk. The same 100 randomly selected audio chunks generated by the TTS models and HiFi-GAN vocoder were transcribed by these ASR systems. We computed the Character Error Rate (CER) by comparing the ASR-generated transcripts with the ground truth transcripts. Additionally, we computed the CER for the ground truth audio to account for the inherent error rates of the ASR models. The results are summarized in Table~\ref{tb:avg-cer}.

\begin{table}[H]
\centering
\caption{\textbf{Average CER of outcomes of TTS models.} \textbf{GT Spec} refers to the utterances
with ground truth spectrograms but HiFi-GAN-synthesized waveforms, and \textbf{GT Waveform} refers to the natural speech samples.}
\label{tb:avg-cer}

\begin{tabular}{lccccc}
\toprule
 & \textbf{Baseline 1 (Vits)} & \textbf{Baseline 2 (Glow)} & \textbf{Ours} & \textbf{GT Spec} & \textbf{GT Waveform} \\
 \bottomrule
\textbf{Google API} & 0.1259	& 0.2325 & 0.0956 & 0.0533 & 0.0482\\
\textbf{Vosk} & 0.2095 &	0.2762	& 0.1506 &	0.1406	& 0.1372\\
\bottomrule
\end{tabular}
\end{table}


\newpage
\section{Supplementary Figures}
\label{sec:supplementary-figures}

\begin{figure}[H]
    \centering
    \includegraphics[width=0.7\linewidth]{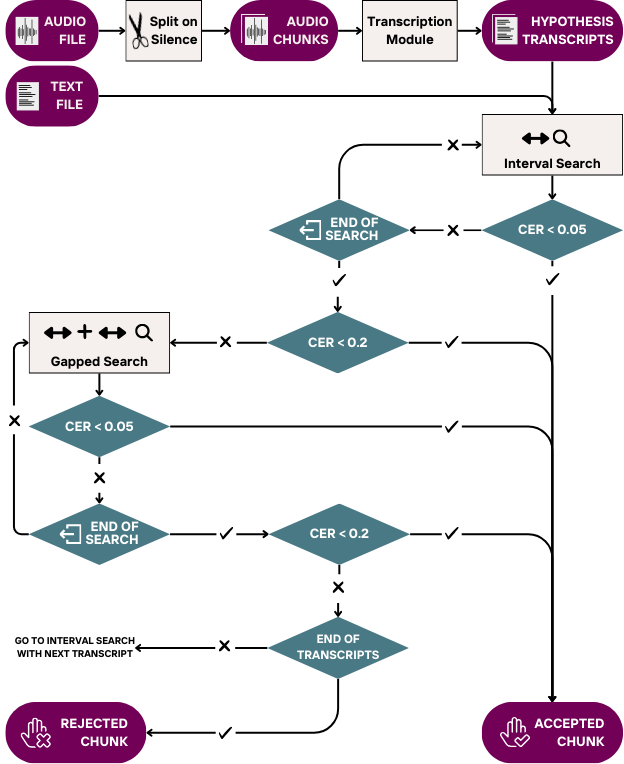}
    \caption{Flowchart of the forced alignment algorithm.}
    \label{fig:appendix-forced-alignment-flowchart}
\end{figure}


\newpage
\section{Supplementary Tables}
\label{sec:supplementary-tables}

\begin{table}[H]
\centering

\caption{The list of tools used in the dataset preparation code; all with open source licenses.}
  \label{tb:appendix-tools}

\begin{tabular}{cccc}
\toprule
\textbf{Tool Name} & \textbf{Usage} & \textbf{Repository Page} & \textbf{License}\\
\midrule

Spleeter \cite{spleeter2020} & \begin{tabular}[c]{@{}c@{}} Source separation \\ (remove background music) \end{tabular} & \href{https://github.com/deezer/spleeter}{GitHub}   & MIT        \\
Parsi.io \cite{parsi.io}  & \begin{tabular}[c]{@{}c@{}} Number extraction \& \\ number to text conversion \end{tabular} & \href{https://github.com/language-ml/parsi.io}{GitHub}  & Apache-2.0 \\
Hazm \cite{roshanai}  & Text normalization                        & \href{https://github.com/roshan-research/hazm}{GitHub}  & MIT        \\
Pydub \cite{pydub} & Silence detection/removal  & \href{https://github.com/jiaaro/pydub}{GitHub}   & MIT        \\
Perpos \cite{bashari2021perpos} & \begin{tabular}[c]{@{}c@{}} Part of speech tagging \\ for sentence tokenization \\  See appendix. \end{tabular} &  \href{https://github.com/mhbashari/perpos}{GitHub} & MIT\\
Vosk \cite{vosk2024} & Forced alignment         & \href{https://github.com/alphacep/vosk}{GitHub}   & Apache-2.0 \\
Whisper-fa \cite{speechbrain2023whisper, speechbrain} & Forced alignment         & \href{https://huggingface.co/speechbrain/asr-whisper-large-v2-commonvoice-fa}{HuggingFace}   & Apache-2.0 \\
Wav2vec2-v3 \cite{m3hrdadfi2021wav2vec2} & Forced alignment         & \href{https://huggingface.co/m3hrdadfi/wav2vec2-large-xlsr-persian-v3}{HuggingFace}  & - \\
Wav2vec2-fa \cite{wav2vec2-xlsr-multilingual-53-fa} & Forced alignment         & \href{https://github.com/Hamtech-ai/wav2vec2-fa}{GitHub}   & Apache-3.0 \\
Hezar \cite{hezar2023} & Forced alignment         & \href{https://github.com/hezarai/hezar}{GitHub}   & Apache-2.0 \\
JiWER  \cite{jiwer} & CER calculation                      & \href{https://github.com/jitsi/jiwer}{GitHub} & Apache-2.0 \\
\bottomrule
\end{tabular}
\end{table}

\end{document}